\documentclass{article}
\usepackage{graphicx} 
\usepackage{tabularx}
\usepackage{multicol}
\usepackage{booktabs}
\usepackage{multirow}
\usepackage{url}

\graphicspath{ {./figs/} }
\usepackage{tikz}
\usepackage[many]{tcolorbox}    	

\title{Cybersecurity Data Extraction from Common Crawl}
\author{Ashim Mahara}
\date{March 2025}

\begin{document}

\maketitle

\section{Introduction}

Generative Large Language Models (LLMs) are pretrained on massively large and diverse sets of text data (\cite{devlin-etal-2019-bert}, \cite{gpt-pretrain}). The datasets used for the pretraining are largely general and multilingual, and contain a diverse set of texts from various sources such as the web, public books\footnote{gutenberg.org}, research papers \footnote{arxiv.org}, Wikipedia \footnote{wikipedia.com} articles and StackExchange \footnote{stackexchange.com} (\cite{C4}, \cite{thepile}, \cite{Slimpajama-dc}, \cite{redpajama}).

Due to the heterogeneous nature of the datasets, the models trained using the datasets perform well for a variety of general downstream natural language tasks such as natural language understanding and reasoning \cite{C4} but recent studies \cite{dontstoppretraining} have shown that the models can benefit from further pre-training on domain-specific or task-specific datasets.

While there are various evaluation datasets and benchmarks for cybersecurity (\cite{secure}, \cite{cybermetric}, \cite{cyberpal.ai}, \cite{cyberq}), there is a general lack of pretraining datasets that can help to further align the models to cybersecurity. To our knowledge, Primus \cite{primus_dataset} is the only publicly available cybersecurity-focused pre-training dataset. 

Thus, we introduce \textbf{Alpha-Root} as a new large text corpus from Common Crawl \footnote{commoncrawl.org} --- intended to be used as a pre-training dataset --- by extracting cybersecurity-focused domains from the webgraph provided by Common Crawl. We utilize the Leiden \cite{leiden} algorithm for community detection to mine the web graph for domains that represent the cybersecurity community in the webgraph. We then extract texts from webpages of the members of the community that are also present in the FineWeb-Edu \cite{fineweb} dataset. We also present evaluations of the trained models - for both Alpha-Root and Primus-FineWeb - on the MMLU:Computer\_Security \cite{hendrycks2021measuringmassivemultitasklanguage} subset of MMLU for evaluations. We show that our dataset consistently matches or outperforms the Primus-FineWeb dataset on several n\_shot evaluations on different floating point precision runs. 
\newtcolorbox{boxA}{
    fontupper = \bf,
    boxrule = 1.5pt,
    colframe = black 
}

\section{Background}

\subsection{Deep Learning}

Deep Learning \cite{deeplearning} is a subset of machine learning, and the broader field of artificial intelligence, that utilizes data to train artificial neural networks capable of performing a variety of tasks. It has revolutionized the technology industry and has been widely adopted by various scientific communities and industries to perform complex processing and analysis of structured and unstructured data. Deep learning has been applied in various domains, from identifying animals or objects in images (general image classification tasks) to detecting algorithmically generated domains (specialized cybersecurity classification tasks) queried by hosts in a computer network \cite{dga_paper}.

Generally, training a deep learning model $M$ on a classification task consists of learning a function $F$ that maps a collection of inputs $X$ to a target output space $Y$. To tune the weights of a neural network, backpropagation \cite{backprop_paper} is used to minimize the difference between the output generated by $M$, usually denoted by $Y'$, and the target output $Y$. A loss function $L$ quantifies the discrepancy between the predicted output $Y'$ and the ground truth label $Y$, guiding the optimization of $M$. In essence, we try to minimize the difference between the probability distributions of the ground truth class $Y$ and the predicted class $Y'$ given $X$.

\begin{boxA}
    Mathematically, the objective loss function for a classification task is usually a slight modification of the cross-entropy loss from the original logistic regression model \cite{logistic_regression} that expands the loss function to handle scenarios where there are more than two classes, such as:

    \begin{equation} \label{eq:1}
        \mathcal{L} = -\frac{1}{N} \sum_{i=1}^{N} \sum_{c=1}^{C} y_{i,c} \log(y'_{i,c})
    \end{equation}
    
    where $N$ denotes the number of samples in the input $X$, $C$ denotes the number of target classes, $y_{i,c}$ is a is a one-hot encoded ground truth indicator for sample $i$ and class $c$, and $y'_{i,c}$ is the predicted probability of sample $i$ belonging to the class $c$ as predicted by the model $M$. 
\end{boxA}

\subsection{Deep Learning in Cybersecurity}

There has been a growing interest of deep learning applications in cybersecurity over the years. (\cite{deeplearning_cyber_survey}) Researchers are finding creative ways to integrate the probabilistic models to different tasks such as Malware Classification \cite{malware_deep_learning}, Phishing Website Detection \cite{phishpedia}, and Anomaly Detection and Log Analytics \cite{log_anomaly_lstm}. 

While a number of different deep learning algorithms such as CNNs \cite{intro_cnn} and LSTMs \cite{lstm_paper} are used in the literature for various tasks \cite{deep_learning_algo_cyber_status}, there has been a meteoric rise in adoption of LLMs in cybersecurity. Zhang et al. \cite{llms_cybersecurity} list various different applications of LLMs in cybersecurity such as Indicator of Compromise (IoC) extraction from Cyber Threat Intelligence (CTI) reports, Source Code Analysis, Fuzzing, and LLM Assisted Red-Teaming among others.


\subsection{Large Language Models}

Large Language Models (LLMs) are a family of deep learning models  --- primarily based on the Transformer architecture by Vaswani et al. \cite{transformers} --- that are used to process sequence of texts. Since deep learning requires numerical representations of the data, the texts are converted to vector representations by \textit{tokenizing} the text and converting the resultant \textit{tokens} to their vector representations for further processing by utilizing an \textit{embedding layer}. \cite{bengio_neural_lm}

\begin{boxA}
Tokenization is the process of creating a dictionary of key-value pairs that maps a token to its corresponding index. An embedding layer then converts the token index to a vector representation for the token. The embeddings --- that are the vector representations of the tokens --- are jointly trained with the language model itself and do not generally need to be trained separately for modern LLMs. However, there is an extra positional encoding layer that adds sequential information to the embeddings that is needed for LLMs utilizing the transformer architecture. \cite{transformers}
\end{boxA}

Transformers utilize the attention mechanism \cite{badhnau_attention} to attribute attention scores to each token in a sequence of text. The attention score determines the importance of a previously seen token while generating the next token in the sequence. They are also pretrained on very large text corpora \cite{gpt-pretrain}.  The pretraining of the language model is performed in an unsupervised learning setting where the training objective consists of minimizing the cross-entropy loss while predicting the next token based on the previous tokens in a sequence of text input. Conversely, it can also be formulated as maximizing the conditional probability of a single token given a sequence of previous tokens.

\begin{boxA}
Radford et al. \cite{gpt-pretrain} formulate the unsupervised pretraining as: \\\\
    Given a sequence of tokens $U = \{u_1,...,u_n\}$, the objective for the unsupervised pretraining is to maximize the following likelihood: 

\begin{equation} \label{eq:log-likelihood}
    L(U) = \sum logP(u_i|u_{i-k},...,u_{i-1}; \theta)
\end{equation}
where $k$ is the context size that determines how many previous tokens are used for the current token prediction, and $P$ is the conditional probability that is modeled by a neural network with parameters $\theta$. The parameters are trained using gradient descent \cite{gradient_descent}; and backpropagation \cite{backprop_paper} --- if the neural network is composed of multiple layers.
\end{boxA}

\section{Related Work} \label{related-work}

\subsection{Pretraining of Language Models}

Our work broadly falls under the umbrella category of data collection for unsupervised training of large language models. Ramachandran et al. \cite{first_unsupervised_pretraining} were the first to \cite{first_unsupervised_pretraining} to demonstrate that language models perform better when pre-trained by pre-training and finetuning an encoder-decoder model on their language translation task. Then, Radford et al. \cite{gpt-pretrain} pre-trained the first decoder-only transformer which was a  117 million parameter transformer-based language model using the BooksCorpus \cite{bookscorpus} dataset; and their resultant model GPT-1 is considered to be the first modern LLM. They used the 7000 books available from the corpus to demonstrate that pre-training the model resulted in a 15 point increase (average) when compared to a model without the pre-training stage.

Later, Devlin et al. \cite{devlin-etal-2019-bert} used the BooksCorpus as well as the English Wikipedia (2,500M words) to pre-train a bidirectional language model (BERT) for learning representations from unlabeled text. BERT used a \textit{masked language modeling} (MLM) approach --- also noted by the authors to be a derivation of Taylor's work in \cite{cloze_task_mlm}  --- for the training as well as prediction of entire sentences instead of just the next tokens. BERT, and its derivatives like RoBERTa \cite{RoBERTa}, were SOTA on many information retreival tasks for years and are still widely used for generating textual representations where efficiency is preferred over accuracy by modern standards.

\subsection{Early Datasets for Pretraining of Language Models}

The first known usage of CommonCrawl for pre-training was by Trinh and Le \cite{first_cc_usage} to train a language model based on the RNN \cite{rnn_architecture} architecture. Subsequently, Radford et al. \cite{gpt-2} used a variety of techniques to clean data obtained from CommonCrawl to create the WebText dataset and train the GPT-2 model. GPT-2 \cite{gpt-2} was trained on text exclusively extracted from CommonCrawl and outperformed every other existing work on various benchmarks in natural language processing at that time while demonstrating that unsupervised training of language models resulted in strong multi-task performance on downstream tasks.

While the WebText \cite{gpt-2} dataset might be the first dataset extracted from CommonCrawl for the sole purpose of pre-training a language model, they did not release the dataset publicly. Instead, Gokaslan et al. \cite{Gokaslan2019OpenWeb} utilized the process presented in \cite{gpt-2} in recreating the dataset, labeled as OpenWebText, and released it to the public. Following the works of \cite{gpt-2} and \cite{Gokaslan2019OpenWeb}, Weznek et al. \cite{ccnet} publicly released the first large scale high-quality dataset purposefully built to train language models from CommonCrawl called CCNET. The methods used by CCNET for the construction of the dataset is still used in later works such as \cite{primus_dataset}.

\subsection{Modern Datasets for Pretraining of LLMs}

The Pile \cite{thepile} can be considered as the first open-source large-scale dataset for large language model pre-training. The Pile consisted of data from various sources, including CommonCrawl, and amounted to a staggering 1.2 TiBs. They also used the GoldMiner \cite{goldminer} algorithm to clean the CommonCrawl web scrapes and concluded that the extra information from the WET archive format was counter-productive for the pre-training.

Next, the C4 \cite{C4} dataset was constructed solely from CommonCrawl. The name \textit{Colossal Clean Crawled Corpus} directly implies the nature of the dataset, which --- at 750 GB --- was used to train the T5 model which was the first study to show that language models can be trained on multiple-tasks at the same time as opposed to training at a single task at a time. Models pre-trained with C4 outperformed other models pre-trained with non-C4 datasets; with the closest competition from datasets derived from CommonCrawl rather than other sources.

More recently, RefinedWeb \cite{refined_web} uses a variety of different quantitative and qualitative measures to extract data from CommonCrawl. They introduced MacroData Refinement (MDR) pipeline for CommonCrawl --- that focused heavily on deduplication and filtering --- and trained the Falcon LLM using the data extracted by using their pipeline. They also introduced URL scoring to filter unwanted URLs for topics such as adult content and gambling. RefinedWeb contained 2.8TB or 600B tokens of text extracted solely from CommonCrawl. 

FineWeb \cite{fineweb} is the first work to use a trained neural network-based classifier to filter content based on their quality. They created the FineWeb-Edu subset from the FineWeb dataset by using a LLM annotator to create training samples for the training the edu-classifier and later utilizing the classifier on the superset of the data. FineWeb-Edu was demonstrated to surpass every other dataset, at that time, for training LLM models when compared against every other publicly available dataset.

\subsection{Cybersecurity Datasets}

While there are numerous works available for general pre-training of LLMs, as outlined previously in this section, there is only one work on cybersecurity-specific pre-training dataset. PRIMUS \cite{primus_dataset} builds upon the work of \cite{fineweb} to build a classifier that filters cybersecurity related topics from FineWeb. They create the PRIMUS-FINEWEB data by combining the data obtained from the previous step, and a combination of various other sources, to create the PRIMUS-PRETRAINING dataset; which is the first cybersecurity-focused pretraining dataset.

Although FineWeb is a leading general-purpose dataset, FineWeb-Edu is an even better version of FineWeb -- introduced in the same paper \cite{fineweb} -- that was curated using a LLM-based classifier that ranked content based on their educational value. PRIMUS did not use FineWeb-Edu to curate their data and instead opted for the FineWeb as their data-source. 

In contrast of FineWeb-Edu ranker, PRIMUS trained a binary classifier that distinguishes whether given content is cybersecurity related or not, and opted to use threshold values from the classifier's output; which is the probability that a certain given text is from cybersecurity domain. 

Our previous previous works such as SECURE \cite{secure} have hinted at using graph methods to filter data-sources from CommonCrawl. We continue on this direction by using Community Detection algorithms - namely the Leiden Algorithm \cite{leiden} - on Common Crawl Web Graph to mine for domains related to cybersecurity. 

\section{Alpha-Root}

\subsection{Overview}

We introduce a novel method for extracting domain-specific web data from CommonCrawl\footnote{commoncrawl.org} by utilizing community detection algorithms. We leverage the Leiden \cite{leiden} algorithm to extract cybersecurity-specific community of web domains from the Common Crawl Web Graph\footnote{As referred to cc-graph throughout the document} by using a list of seed domains. The extracted web-domains are subsequently used to aggregate webpages from FineWeb-Edu. We used this approach to create a cybersecurity pretraining text corpus called \textbf{Alpha-Root}. Alpha-Root is an early cybersecurity focused pre-training dataset consisting of \textbf{3 billion} tokens sourced from 3.3 million webpages.

\begin{figure}[h]
    \centering
    \includegraphics[width=\textwidth]{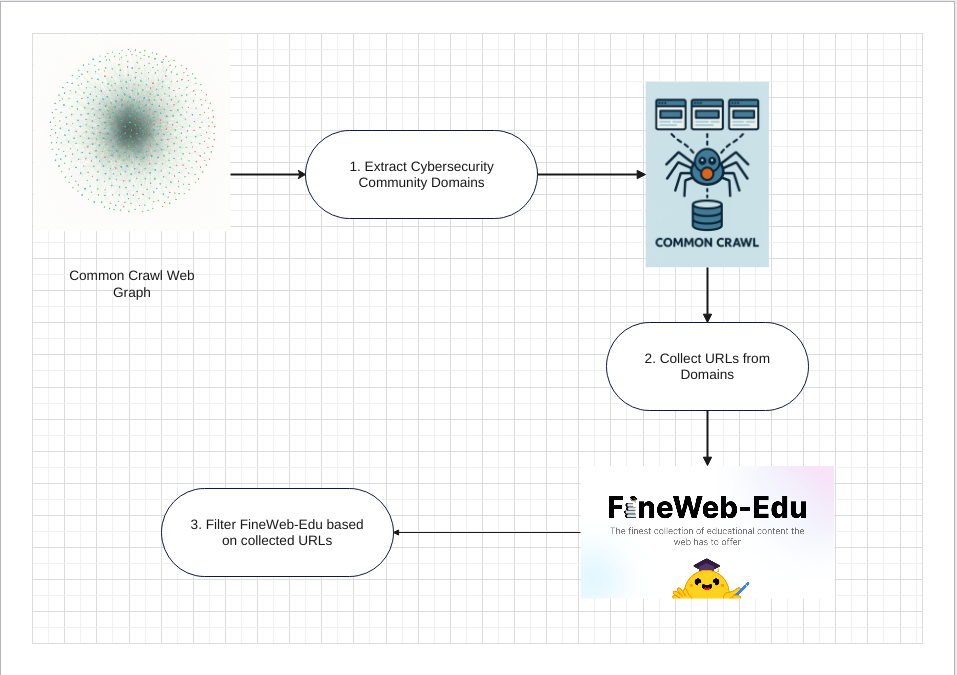}
    \caption{Dataset Extraction Process}
    \label{fig:process}
\end{figure}

\subsection{Cybersecurity Community Detection in Web Graph}

\subsubsection{Web Graph}

CommonCrawl release web graphs for each of their crawls. The cc-graph is a graph of nodes (domains) and edges between the nodes when the nodes are hyperlinked through a webpage. Essentially, there exists an edge between two domains if there is at-least one hyperlink between webpages of the domains.

The cc-graph includes more than \textbf{100 million} nodes and more than \textbf{1.8 billion} edges. The processing of such a large graph presents several challenges on its own since commoncrawl only provides the data in an edge-list format while a lot of existing software libraries require an adjacency list to work on the graph.

We used a memory-mapped sparse matrix in compressed sparse row format to compute and store the adjacency list of the graph. The adjacency list allowed us to make use of several available libraries such as Cugraph \footnote{https://github.com/rapidsai/cugraph} and Scikit-Network \footnote{https://github.com/sknetwork-team/scikit-network} to run the community detection algorithm on modern GPUs - which vastly speeds up the process while extracting communities from the web graph.

\subsubsection{Extraction of Cybersecurity Community}

While the topic of community detection has been well studied in the past by pioneers in the field of networks such as Newman et al. in their works(\cite{Newman_2004}, \cite{Newman_2006}), the research on the topic of community detection in the web has stagnated in recent times. 

The most prominent research in the field was performed more than two decades ago in works such as \cite{Flake_Lawrence_Giles_Coetzee_2002}, \cite{Gibson_Kleinberg_Raghavan_1998}, and \cite{Kumar_Raghavan_Rajagopalan_Tomkins_1999}; with majority of the works opting for some variant of hyperlink induced community detection. While the efficacy of the method cannot be understated for data collection - as evident by the success of the DeepSeekMath \cite{deepseekmath} model in which the researchers crawled hyperlinks from a seed of curated webpages to collect their dataset - such methods can be considered primitive and inefficient since the introduction of modern methods of community detection such as Modularity Maximization \cite{Newman_2006}. Even works such as PRIMUS \cite{primus_dataset} follow the steps of FineWeb \cite{fineweb} in training a classifier that is used to filter cybersecurity related web pages but this approach entails processing web-scale amount of text through the classifier which may not be feasible for adoption by resource-constrained audience. 

In contrast, we utilized the Leiden algorithm \cite{leiden} - an algorithm based on modularity maximization - to extract a collection of domains belonging to the cybersecurity community within the web graph. 

\subsection{Dataset Curation}

Once the domain were extracted from the web graph, we used the common crawl index to further extract webpages from the domains. We collected \textbf{70 million} unique webpages across all of the crawls from common crawl. While we wanted to download all of the webpages that we collected and run a classifier on the content for further filtering, it seemed infeasible to do so due to the download rate limits from Amazon Web Services (AWS)\footnote{https://aws.amazon.com/s3/} and due to our computing resources limitations.

We instead opted to filter the FineWeb-Edu dataset based on our collection of URLs. This approach allowed us to collect the sources for the data at large without explicitly filtering based on the content. We were able to curate our Alpha-Root dataset in 8 hours; with a couple of restarts since the FineWeb-Edu (5 TiB) dataset was streamed and filtered in memory due to storage constraints on our computing cluster. The resultant dataset contains \textbf{2.8 million} URLs from the original \textbf{72.8 million} URLs that we collected. 

\begin{table}[h!]
\centering
\newcolumntype{Y}{>{\centering\arraybackslash}X} 
\begin{tabularx}{\textwidth}{lYY}
\toprule
\textbf{Dataset}          & \textbf{Number of Examples} & \textbf{Number of Tokens} \\
\midrule
Primus‑FineWeb            & 3\,386\,733                  & 2.57B              \\
Alpha‑Root                & 2\,866\,811                  & 3B                  \\
Alpha-Root-Dedup          & 684\,016                     & 714M                \\
\bottomrule
\end{tabularx}
\caption{Comparison of dataset sizes for Primus‑FineWeb and Alpha‑Root.}
\label{tab:dataset-comparison}
\end{table}

\section{Training and Evaluation}

\subsection{Training}

The models were trained using the hyperparameters detailed in Table~\ref{tab:hyperparams}. We combined efficient quantization techniques, Low-Rank Adaptation (LoRA), and large-scale distributed training for efficient training on our limited resources. Below, we contextualize key design choices and their implications.

\subsubsection{Computational Efficiency}
To optimize memory usage and throughput, the training employed:
\begin{itemize}
    \item \textbf{4-bit Quantization} (NF4 type) with \texttt{bfloat16} compute dtype, reducing memory footprint while preserving gradient precision \cite{dettmers2023qloraefficientfinetuningquantized}.
    \item \textbf{8-bit AdamW} for optimizer states, further lowering memory requirements.
    \item \textbf{Gradient Accumulation} (8 steps) to simulate larger batch sizes ($8 \times 8 = 64$ effective batch size per device) without increasing memory pressure.
\end{itemize}

\subsubsection{Architecture \& Adaptation}
The base model (1.8B parameters) was adapted via LoRA to balance parameter efficiency and task-specific learning:
\begin{itemize}
    \item LoRA rank $r=128$ with $\alpha=256$ applied to attention and feed-forward projections.
    \item Only 346M (16.8\% of total) parameters were trainable, focusing adaptation on critical modules (\texttt{lm\_head}, \texttt{embed\_tokens}).
    \item A sequence length of 8192 tokens to enable long-context processing due to the datasets containing texts upto 140k tokens (PRIMUS); and maximum of 120k tokens for Alpha-Root.
\end{itemize}

\subsubsection{Distributed Training}
Training was conducted on \textbf{two NVIDIA GH200 nodes}, leveraging their unified CPU-GPU memory. Each GH200 node contains 95GB of VRAM available for training; we were able to maintain 80-85\% effective utilization for the training runs. The 1-epoch training schedule (cosine LR, 5e-4 peak) prioritized rapid convergence, with a 1\% warmup ratio to stabilize early optimization.

\begin{table}[h!]
\centering
\small
\begin{tabularx}{\textwidth}{l l X}
\toprule
\textbf{Category}           & \textbf{Parameter}                  & \textbf{Value}                                                   \\
\midrule
\multirow{5}{*}{Precision \& Quant.}
                            & bf16                                & true                                                             \\
                            & load\_in\_4bit                      & true                                                             \\
                            & bnb\_4bit\_quant\_type              & nf4                                                              \\
                            & bnb\_4bit\_compute\_dtype           & bfloat16                                                        \\
                            & bnb\_4bit\_quant\_storage           & uint8                                                            \\
\midrule
\multirow{5}{*}{Architecture}
                            & hidden\_size                        & 2048                                                             \\
                            & intermediate\_size                  & 8192                                                             \\
                            & num\_hidden\_layers                 & 24                                                               \\
                            & num\_attention\_heads               & 32                                                               \\
                            & max\_seq\_length                    & 8192                                                             \\
\midrule
\multirow{4}{*}{Optimizer \& LR}
                            & optim                               & adamw\_8bit                                                     \\
                            & learning\_rate                      & 5e-4                                                             \\
                            & lr\_scheduler\_type                 & cosine                                                           \\
                            & warmup\_ratio                       & 0.01                                                             \\
\midrule
\multirow{3}{*}{Training Loop}
                            & num\_train\_epochs                  & 1                                                                \\
                            & per\_device\_train\_batch\_size     & 8                                                                \\
                            & gradient\_accumulation\_steps       & 8                                                                \\
\midrule
\multirow{5}{*}{LoRA \& PEFT}
                            & rank (r)                            & 128                                                              \\
                            & lora\_alpha                         & 256                                                              \\
                            & lora\_dropout                       & 0.05                                                             \\
                            & target\_modules                     & q\_proj, k\_proj, v\_proj, up\_proj, down\_proj, gate\_proj, o\_proj \\
                            & modules\_to\_save                   & lm\_head, embed\_tokens                                         \\
\midrule
\multirow{2}{*}{Model Parameters}
                            & total                               & 2\,057\,406\,464                                                    \\
                            & trainable                           & 346\,030\,080                                                       \\

\bottomrule
\end{tabularx}
\caption{Key hyper‑parameters for training SmolLM‑1.7B with LoRA adapters}
\label{tab:hyperparams}
\end{table}

\subsection{Evaluation} \label{ref:evaluation}

We used the lm-eval \cite{eval-harness} library for our evaluations. 

\subsubsection{MMLU: Massive Multitask Language Understanding}
The \textbf{Massive Multitask Language Understanding (MMLU)} benchmark \cite{hendrycks2021measuringmassivemultitasklanguage} is a comprehensive evaluation framework designed to assess the knowledge and reasoning capabilities of machine learning models across diverse domains. It consists of multiple-choice questions spanning 57 subjects, including STEM, humanities, social sciences, and professional disciplines. MMLU evaluates both \textit{factual knowledge} and \textit{conceptual understanding}, making it a robust measure of a model's generalization ability and multitask proficiency. We specifically chose MMLU for our evaluation due to the benchmark's compatibility with models that have not been instruct-finetuned.

A specialized subset of MMLU, \textbf{MMLU\_COMPUTER\_SECURITY}, focuses on assessing a model's understanding of cybersecurity principles. This subset includes questions on topics such as:
\begin{itemize}
    \item Cryptographic protocols (e.g., AES, RSA),
    \item Network security (e.g., firewalls, intrusion detection),
    \item Software vulnerabilities (e.g., buffer overflows, SQL injection),
    \item Access control models (e.g., RBAC, MAC).
\end{itemize}
Performance on this subset indicates a model's ability to reason about technical and theoretical aspects of computer security, which is critical for applications in automated threat analysis, secure code generation, and adversarial robustness. 

\section{Experiment Results}

As explained in \ref{ref:evaluation}, we opted to use MMLU:Computer\_Security as our evaluation metric due to the log likelihood \cite{hendrycks2021measuringmassivemultitasklanguage} calculations for the metrics. We performed 5 n\_shot evaluations for the models with 2 sets of floating point precision - this decision was made due to the usage of lower quantization in the training phase since we wanted to perform fair evaluations of the trained and the base models.

We compared our trained models against the base SmolLM-1.7B model as well as the pretrained SmolLM model later continually pre-trained on Primus-FineWeb. We notice that our models, whether pretrained with a higher batch size or with the same batch size as the Primus-FineWeb, consistently performed on the same level as the Primus-FineWeb. The \textbf{checkpoint-5700} was included to assess whether a lower amount of training steps - coupled with a dataset with duplicates in case of Alpha-Root - would skew the evaluations. 

We notice irregularities in the few shot evaluations for both floating point variations of the evaluations. However, we also note that our dataset consistently performs on the same level as Primus-FineWeb -  and even better in a lot of the scenarios - while requiring lesser computational resources for the extraction of the corpus. We also performed the evaluations with different seeds for a fairer comparison but it resulted in the same scores for all of the experiments we ran so we left it as \textbf{1234}. We used the \textit{lm-eval} \cite{eval-harness} library from EleutherAI\footnote{https://github.com/EleutherAI/lm-evaluation-harness} for generating the evaluation scores.

\begin{table}[h!]
\centering
\newcolumntype{Y}{>{\centering\arraybackslash}X}
\begin{tabularx}{\textwidth}{lYYYYY}
\toprule
\textbf{Model} & \multicolumn{5}{c}{\textbf{mmlu\_computer\_security}} \\
\cmidrule(r){2-6}
 & \textbf{0-shot} & \textbf{1-shot} & \textbf{2-shot} & \textbf{3-shot} & \textbf{5-shot} \\
\midrule
SmolLM-1.7B (Base)                  & 0.28            & 0.32            & \textbf{0.37}   & 0.26            & \textbf{0.33}   \\
Primus-FineWeb                      & \textbf{0.32}   & \underline{0.33}& 0.29            & \underline{0.31}& 0.28            \\
Alpha-Root (256bsz, Full Training) & 0.26            & \textbf{0.35}   & 0.27            & \textbf{0.32}   & \underline{0.30}\\
Alpha-Root (checkpoint-5700)        & \underline{0.30}& 0.26            & 0.30 & 0.28            & 0.29            \\
\textbf{Alpha-Root (Full Training)} & 0.25           & 0.32           & \underline{0.31}           & 0.28           & 0.29           \\
\bottomrule
\end{tabularx}
\caption{Performance on the \texttt{mmlu\_computer\_security} benchmark with 0‑, 1‑, 2‑, 3‑, and 5‑shot evaluations. BF16 inference. Highest scores are \textbf{bolded}, second‑highest are \underline{underlined}.}
\label{tab:mmlu-computer-security-bf16}
\end{table}

\begin{table}[h!]
\centering
\newcolumntype{Y}{>{\centering\arraybackslash}X}
\begin{tabularx}{\textwidth}{lYYYYY}
\toprule
\textbf{Model} & \multicolumn{5}{c}{\textbf{mmlu\_computer\_security}} \\
\cmidrule(r){2-6}
 & \textbf{0-shot} & \textbf{1-shot} & \textbf{2-shot} & \textbf{3-shot} & \textbf{5-shot} \\
\midrule
SmolLM-1.7B (Base)                  & \underline{0.27}& 0.34            & \textbf{0.32}   & \textbf{0.30}   & \textbf{0.31}   \\
Primus-FineWeb                      & \textbf{0.31}   & \underline{0.35}& 0.28            & \textbf{0.30}   & \underline{0.29}\\
Alpha-Root (256bsz, Full Training) & 0.26            & \textbf{0.38}   & 0.28            & 0.26            & \underline{0.29}\\
Alpha-Root (checkpoint-5700)        & \textbf{0.31}   & 0.27            & \underline{0.30}& \underline{0.28}& 0.28            \\
\textbf{Alpha-Root (Full Training)} & 0.26           & 0.34           & \textbf{0.32}           & \textbf{0.30}           & \underline{0.29}           \\
\bottomrule
\end{tabularx}
\caption{Performance on the \texttt{mmlu\_computer\_security} benchmark with 0‑, 1‑, 2‑, 3‑, and 5‑shot evaluations. FP32 inference. Highest scores are \textbf{bolded}, second‑highest are \underline{underlined}.}
\label{tab:mmlu-computer-security-fp32}
\end{table}

\section{Discussion}

In this section, we are going to share our findings in data collection and pre-training:

\subsection{Data Collection at Scale is Hard}

When we initially started to collect the domains by leveraging the web graph, we had to find workarounds for managing the memory and storage requirements since creating an adjacency list out of a large graph like the commoncrawl web graph -- with $>$100 million nodes and $>$1.8 billion edges -- would require petabyte scale of memory just for storage. We found that using memory mapped file for the sparse matrix and splitting the original edge list into several small files for processing the small \textit{splits} was sufficient. We also found the UNIX program \textit{split} to be sufficient for the task while the splits themselves were processed using a custom Python script.

The memory and storage impediments were persistent during the text extraction phase too. Due to a non-trivial amount of web pages (~70M) to extract from CommonCrawl, we opted to use FineWeb-Edu as our data source. We faced 3 hurdles during this phase,

\begin{itemize}
    \item \textbf{Downloading Web Pages}: The webpages from commoncrawl were hosted in AWS S3 which throttled our requests to 30 iterations per second; this made it impossible to download the web pages in a reasonable time - with the estimated time being 30-36 days for downloading the WARC records alone.

    \item \textbf{Scoring the text from Web Pages}: We had initially planned to score the texts by fine-tuning a scoring model trained on a FineWeb-Edu inspired LLM-as-a-Judge dataset for filtering out high quality cybersecurity content. While the training of the model was completed, the resultant model could not be used for scoring the texts due to computational restraints. 
\end{itemize}

\subsection{Common Crawl is Massive}

During our data collection phase, we noticed that several websites now disallow the common crawl bots from crawling their websites.\footnote{https://www.bleepingcomputer.com/robots.txt} Even with the restrictions, each individual release of the official crawls contain terabytes of data - the latest April 2025 crawl contains 468 TiB of data. \footnote{https://commoncrawl.org/blog/april-2025-crawl-archive-now-available}

The scale of the available data makes it infeasible for domain-specific model authors to naively collect training data using scoring based models which requires the processing of every example of text in the crawls. Alternatively, our method provides a leaner approach to the problem by directly finding good domains for crawling and text extraction by exploiting the graph features of the commoncrawl web graph. 

We found that out of the \textbf{15240} unique domains present in Alpha-Root, \textbf{9250} are also present in the PRIMUS dataset. This intersection of common data sources also signals that our method is able to mine sources of good data without directly inferring their relevance from processing their content - which is both computationally expensive and time consuming. 

\subsection{Text Scoring Models are Good Too}

While our method focuses on extracting domains and web pages from those domains using the web graph, the extracted web pages can undoubtedly be filtered even further by using a scoring model. The text-scoring model can be integrated into our method after the URLs have been collected from the domains in the mined cybersecurity community. 
\section{Future Work}

There are a few unfinished tasks directly related to our work that we would like to highlight as future work. We list some of them here as:

\begin{itemize}
    \item \textbf{Evaluate the domain extraction phase.} In this paper we just used the extracted domains and constructed our dataset, for a deeper analysis we could extract the domains at multiple configurations of the modularity maximization algorithm.
    \item \textbf{Train on the de-duplicated Alpha-Root dataset.} We haven't had the chance to perform the training and evaluations on the de-duplicated dataset.  
    \item \textbf{Download and filter the original 70 million URLs links using the scoring model.} Although this would require a lot of processing and storage, this would be an interesting direction to pursue.
    \item \textbf{Continue Pretraining without using LoRA adapters.} We can see that there are some fluctuations on the MMLU evaluations based on the floating point type used during inference. It would be a natural direction for us to continue pretrain in a higher precision for more rigorous comparisons.
\end{itemize}

\section*{Acknowledgments}

Part of the work was completed during Summer 2024 under tutelage of Prof. Nidhi Rastogi, and Spring 2024 Complex Networks class by Prof. Nishant Malik. Current dataset work was completed as part of Deep Learning Security class by Prof. Matthew Wright. In loving remembrance of \url{https://github.com/rootnp} (alpha-root).
\bibliographystyle{plain}
\bibliography{references}

\end{document}